\documentclass[
    ,final            
  ]
  {aipproc}

\layoutstyle{6x9}

\def\Msun{M$_{\odot}$}

\begin{document}

\title{Echo Tomography of Sco X-1 using Bowen Fluorescence Lines}


\author{J. Casares}{
  address={Instituto de Astrof\'\i{}sica de Canarias, 38200 La Laguna, 
  Tenerife, Spain}
}
\author{T. Mu\~noz-Darias}{
  address={Instituto de Astrof\'\i{}sica de Canarias, 38200 La Laguna, 
  Tenerife, Spain}
}
\author{I.G. Mart\'\i{}nez-Pais}{
  address={Instituto de Astrof\'\i{}sica de Canarias, 38200 La Laguna, 
  Tenerife, Spain}
}
\author{R. Cornelisse}{
  address={School of Physics \& Astronomy, Univ. of Southampton, Southampton  
  SOB17 1BJ, UK}
}
\author{P.A. Charles}{
  address={School of Physics \& Astronomy, Univ. of Southampton, Southampton  
  SOB17 1BJ, UK}
}
 
\author{T.R. Marsh}{
  address={Dept. of Physics, Univ. of Warwick, Coventry CV4 7AL, UK
}
}

\author{V.S. Dhillon}{
  address={Dept. of Physics \& Astronomy, Univ. of Sheffield, Sheffield  
  S3 7RH, UK
}
}

\author{D. Steeghs}{
  address={Harvard-Smithsonian Center for Astrophysics, Cambridge  
  MA 02138, USA
}
}

\begin{abstract}
We present preliminary results of a simultaneous X-ray/optical 
campaign of the prototypical LMXB Sco X-1 at 1-10 Hz time 
resolution. Lightcurves of the high excitation Bowen/HeII emission lines 
were obtained through narrow interference filters with ULTRACAM, and 
these were cross-correlated with X-ray lightcurves. We find evidence 
for correlated variability, in particular when Sco X-1 enters the Flaring 
Branch. The Bowen/HeII lightcurves lag the X-ray lightcurves with a light 
travel time which is consistent with reprocessing in the companion star.  
 
\end{abstract}

\maketitle


\section{Introduction:Irradiation in LMXBs}

Optical emission in persistent low mass X-ray binaries (hereafter LMXBs) is 
triggered by reprocessing of the powerful, almost Eddington limited, X-ray 
luminosity ($L_{\rm x}\simeq 10^{38}$ erg s$^{-1}$) in the gas around the 
compact object. This is supported by independent arguments such as 
(i) the statistical distribution of dereddened $(U-B)$, $(B-V)$ colours 
(or $F_{\nu} \sim$ const., see \cite{van95}) which can be accounted for by 
redistribution of 
high energies into UV+optical through irradiation models (e.g. \cite{vrt90}); 
(ii) the detection of optical counterparts of Type I X-ray bursts, with delay 
times consistent with binary separations (e.g. \cite{grind74}); 
(iii) the presence of the broad emission feature at $\lambda\lambda$4640-50 
associated with a blend of CIII/NIII/OII powered by X-ray photoionization and 
Bowen fluorescence emission (hence refered to as {\it the Bowen blend}; 
\cite{mcc75}, \cite{scha89});
(iv) suppression of outburst cycles caused by irradiation-induced heating of 
the outer disc \cite{van96}, \cite{king96}.  

The accretion disc subtends the largest solid angle as viewed by the
X-ray source, and is therefore responsible for the majority of the
irradiation component.  The spectroscopic features of the weak
companion star, on the other hand, are completely swamped by the
disc's reprocessed light, with the exception of a few long-period
LMXBs with evolved companions such as Cyg X-2
(\cite{casa98}). Therefore, dynamical studies have classically been
restricted to the analysis of X-ray transients during quiescence
(e.g. see \cite{char04}).

\subsection{Fluorescence Emission from Donor Stars}

However, this situation has changed recently thanks to the discovery
of narrow emission components arising from the donor star in Sco X-1
\cite{stee02}.  High resolution spectroscopy, obtained with ISIS at
the WHT, revealed many narrow high-excitation emission lines, the most
prominent associated with NIII $\lambda\lambda$4634-41 and CIII
$\lambda\lambda$4647-50 at the core of the broad Bowen blend
(Fig. 1). The NIII lines are powered by fluorescence resonance through
cascade recombination which initially requires seed photons of HeII
Ly${\alpha}$.  These narrow components are not resolved (i.e. their
FWHM is the instrumental resolution) and they move in antiphase with
respect to the wings of the HeII $\lambda$4686 line, which
approximately trace the motion of the compact star. Both properties
(narrowness and phase offset) imply that these components originate in
the irradiated face of the donor star. This work represents the first
detection of the companion star in Sco X-1 and opens a new window for
extracting dynamical information and thereby deriving mass functions in a
population of $\sim$ 20 LMXBs with established optical counterparts.

\begin{figure}
  \includegraphics[height=.45\textheight, angle=-90]{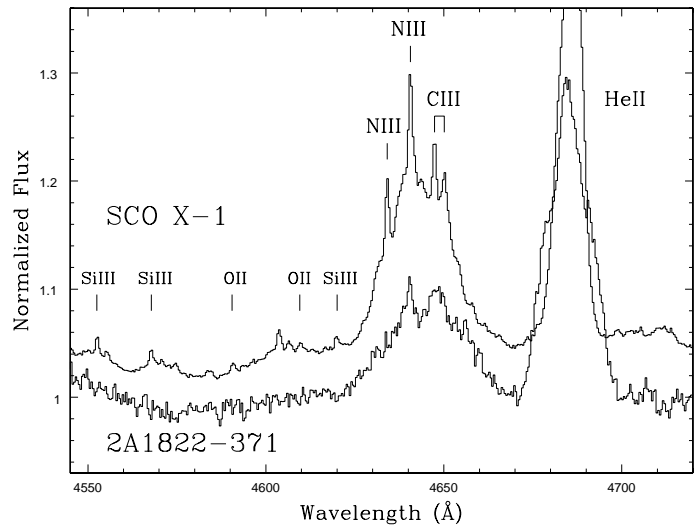}
  \caption{Summed spectra of Sco X-1 (top) and 2A1822-371 (bottom) in the 
  rest frame of the companion star. Adapted from \cite{stee02} and 
  \cite{casa03}.}
\end{figure}

We now know that this property is not peculiar to Sco X-1 but is a feature  
 of persistent LMXBs, as demonstrated by the following examples: 

(i) Radial velocities of narrow Bowen lines in the black hole candidate 
GX339-4, detected during the 2002 outburst, led to a mass function in 
excess of 5.8 \Msun~ and hence provided the first dynamical proof for a black hole
\cite{hynes03}. 

(ii) Velocity information from the Bowen NIII $\lambda$4640 line in the 
eclipsing ADC ({\it accretion disc corona}) pulsar 2A 1822-371, established a 
lower limit to the neutron star's mass of 1.14 \Msun. Moreover, the radial 
velocity curve of the NIII emission is perfectly consistent with the donor's
phase, as expected from the pulse time delay of the 0.59s spin period of the 
neutron star \cite{casa03}. 

(iii) Sharp NIII $\lambda$4640 Bowen emission has been detected in 4U 1636-536, 
4U 1735-444 and the transient millisecond pulsar XTE J1814-338, which lead to 
donor velocity semi-amplitudes in the range 200-300 km s$^{-1}$ 
\cite{casa04}.

\subsection{Echo-Tomography}

One of the most exciting prospects for this new technique is the possibility 
to perform echo-tomography using the Bowen lines. 
Echo-tomography is an indirect imaging technique which uses time delays 
between X-ray and UV/optical lightcurves as a function of orbital phase 
in order to map the reprocessing sites in a 
binary \cite{obrien02}. The optical lightcurve can be simulated by the 
convolution of the (source) X-ray lightcurve with a transfer function which 
encodes information about the geometry and visibility of the reprocessing 
regions. The transfer function quantifies the binary response to the 
irradiated flux as a function of the lag time and it has two main components, 
the accretion disc and the donor star. The latter is strongly dependent on 
the inclination angle, binary separation and mass ratio and, therefore, 
can be used to set tight contraints on these fundamental parameters.  
Successful echo-tomography experiments have been performed on several X-ray 
active LMXBs using X-ray and broad-band UV/optical lightcurves. The results 
indicate that the reprocessing flux is mostly dominated by the large
contribution of the accretion disc (e.g. \cite{hynes98}, \cite{obrien01},
\cite{hynes05}). 

Exploiting emission-line reprocessing rather than broad-band
photometry has two potential benefits: a) it amplifies the response of
the donor's contribution by suppressing most of the background
continuum light (which is associated with the disc); b) since the
reprocessing time in the lines is instantaneous, the response is
sharper (i.e. only smeared by geometry) and also the transfer function
is easier to compute (see \cite{teo05}).  Therefore, we decided to
undertake a simultaneous X-ray/optical campaign on the prototypical LMXB
Sco X-1 with the aim of performing echo-tomography so as to search for
the reprocessed signatures of the donor using Bowen/HeII lines.  As a
first step, here we present our preliminary cross-correlation analysis
which provides evidence for delayed echoes consistent with
reprocessing in the companion star.


\section{Observations}

Simultaneous X-ray and optical data of Sco X-1 were obtained on the nights 
of 17-19 May 2004.  The full 18.9 hr orbital period was covered 
in 12 snapshots, yielding 
20.1 ks of X-ray data with the RXTE PCA.
Only 2 PCA detectors ($2$ and $5$) were used and the pointing offset was 
set to 0$^{\circ}$.71 due to the brightness of Sco X-1. The data were analysed 
using the FTOOLS software and the times corrected to the solar barycenter. 
The STANDARD-2 mode data, with a time resolution of 16s, were used to produce 
a colour-colour diagram which showed that Sco X-1 was in the Normal Branch on
17 and 19 May and in the Flaring Branch on 18 May.
The STANDARD-1 mode, with
a time resolution of 0.125s, was used for the variability analysis.

\begin{table}
\begin{tabular}{lccccr}
\hline
    \tablehead{1}{c}{b}{Date\\} 
  & \tablehead{1}{c}{b}{Exp. time\\(secs)}
  & \tablehead{1}{c}{b}{Seeing\\}
  & \tablehead{1}{c}{b}{Orbital Phases\tablenote{Computed using ephemeris from
  \cite{stee02}}\\}
  & \tablehead{1}{c}{b}{Number of\\XTE windows}
  & \tablehead{1}{c}{b}{X-ray State\\}\\
\hline
17 May 2004 & 0.1 & < 1" & 0.07-0.35 & 4 & Normal Branch\\
18 May 2004 & 0.25-1 & 1"-5" & 0.34-0.73 & 5 & Flaring Branch\\
19 May 2004 & 0.3 & 1"-2"& 0.55-0.95 & 5 & Normal Branch\\
\hline
\end{tabular}
\caption{Observing log}
\label{tab:a}
\end{table}

The optical data were obtained with ULTRACAM on the 4.2m WHT at La Palma.
ULTRACAM is a triple-beam CCD camera which uses two dichroics to split the 
light into 3 spectral ranges: Blue (<$\lambda$3900), 
Green($\lambda\lambda$3900-5400) and Red (>$\lambda$5400). 
It uses frame transfer 1024x1024 Marconi CCDs which are 
continuously read out, and are capable of time resolution down to 500 Hz by 
reading 
only small selected windows (see \cite{dhi01} for details).
ULTRACAM is equipped with a standard set of $ugriz$ Sloan filters. However, 
since we want to 
amplify the reprocessed signal from the companion, we decided to use two narrow 
(FWHM =100 \AA) interference filters in the Green and Red channels, 
centered at $\lambda_{\rm eff}$=4660\AA~ and $\lambda_{\rm eff}$=6000\AA. These 
will block out most of the continuum light and allow us to integrate two 
selected spectral regions: the Bowen/HeII blend and a 
featureless continuum, from which continuum-subtracted lightcurves of the 
high excitation lines can be derived. 
The images were reduced in the standard way with bias subtraction 
and flatfielding. Star counts were extracted using an optimal extraction
algorithm \cite{nay98} and lightcurves were obtained relative to a comparison 
star which is 96 arcsecs NW of Sco X-1. Lightcurves of the Bowen/HeII lines 
were
computed by subtracting the Red (continuum) and Green channel lightcurves.
The seeing was 1-1.5 arcsecs most of the time, 
except for the first two RXTE windows of the second night when it rose to over  
3 arcsecs. Optical observations during the first 3 RXTE visits on 19 May 
were not possible  because of clouds. 
The exposure time was initially set to 0.1s but was increased to 0.25s,
because of weather conditions.  Integrations of 1s were used for the first
window on 18 May, when the seeing was worst. An observing log is presented in
Table 1.  

\begin{figure}[ht]
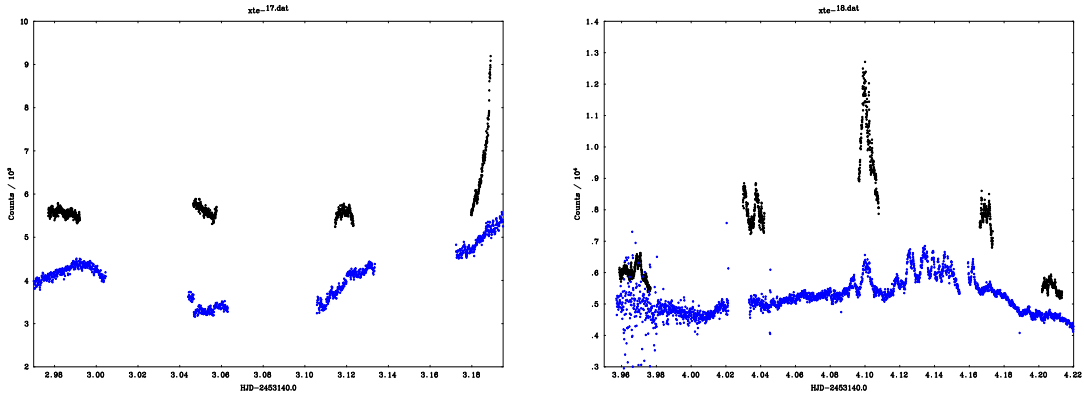

\begin{picture}(250,160)(50,30)
\put(0,0){\includegraphics{xte_uc_17may_all.ps}}
\put(0,0){\includegraphics{xte_uc_18may_all.ps}}
\noindent
\end{picture}
\caption{X-ray and Bowen/HeII lightcurves for the nights of 17 
May (left panel) and 18 May (right panel). The top lightcurves correspond to 
the RXTE data and have been averaged in 4s bins. The bottom lightcurves 
represent the (continuum-subtracted) Bowen/HeII emission lightcurves and have 
been averaged in 8s bins. The Bowen/HeII counts are 
relative to the comparison star and have been multiplied by a factor 23500 for 
display clarity.
}
\label{fig2}
\end{figure}

\section{Results}

Figure 2 presents the X-ray and Bowen/HeII lightcurves corresponding
to the nights of 17 and 18 May. Sco X-1 was at the bottom of the
Normal Branch during most of the first night but moved towards the
Flaring Branch in the last RXTE visit, showing a 50 percent increase
in flux. The amplitude of the X-ray variability is less than 1\% for
the first three windows and no clear correlation with Bowen/HeII is
evident on long timescales.  On 18 May Sco X-1 was in the Flaring
Branch and exhibited large amplitude variability, with large flares
similar to that seen during the third RXTE visit.  The left panel in
Fig. 3 presents a 10 min segment of the third RXTE window, with each
tickmark corresponding to 43s. We note significant X-ray variability,
at the 10 percent level, and some correlated structures can be
identified in the Bowen/HeII lightcurve towards the end of this
interval. This window is centered at orbital phase 0.53 i.e. near the
superior conjunction of the donor star, when the irradiated face of
the donor presents the largest visibility and the light-travel delay
is expected to be at a maximum (around 12s for the binary parameters in
\cite{stee02}). We have calculated cross-correlation functions \cite{gask87} 
for several time intervals within this window, after subtracting a low-order
polynomial fit to the lightcurves in order to remove low-frequency
variations. The right panel in Fig. 3 presents one of the
cross-correlation functions, and shows a clear peak centered at a lag of $\sim$
10-15s. This is in good agreement with the expected delay time for
reprocessing in the companion star at this particular orbital
phase. We have also detected correlated variability in other windows
and we expect to be able to combine this information in order to set
constraints on the binary parameters of Sco X-1.

\begin{figure}[ht]
\begin{picture}(250,160)(50,30)
\put(0,0){\includegraphics{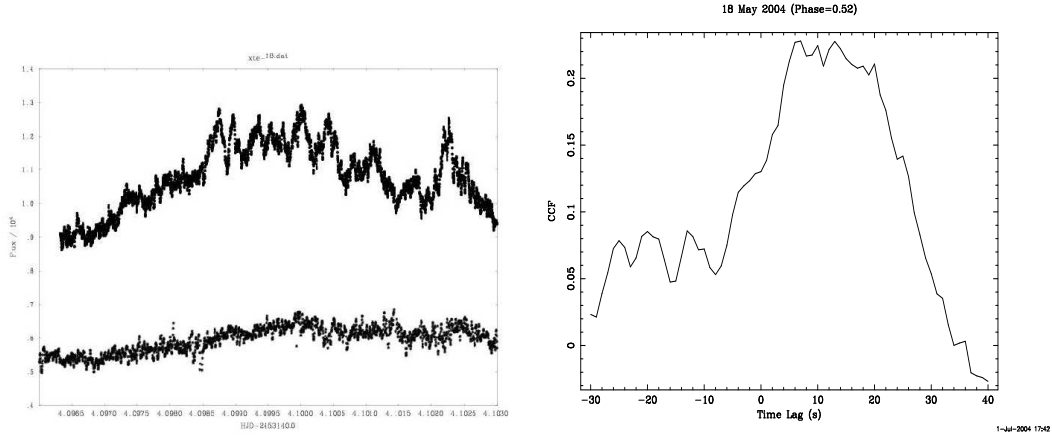}}
\put(0,0){\includegraphics{ccf_18may_w2.ps}}
\noindent
\end{picture}
\caption{Left panel: 10 min detail from the third RXTE visit during 18 May. 
Top lightcurve represents the RXTE data whereas the bottom lightcurve is the 
Bowen/HeII data. Right panel: Cross-correlation function of the two lightcurves 
for the last 3 min interval.}
\label{fig3}
\end{figure}

\section{Summary}
We have obtained simultaneous X-ray/optical photometry of Sco X-1 at 1-10 Hz 
time resolution using RXTE and WHT+ULTRACAM. 
The use of narrow interference filters in ULTRACAM has provided us with 
lightcurves of the Bowen blend + HeII $\lambda$4686 emission lines. Our
preliminary analysis shows evidence for correlated variability, with the 
Bowen/HeII lightcurves lagging the X-rays. The observed 
time delays are consistent with reprocessing in the companion star
and the correlations are most evident when Sco X-1 is in the Flaring 
Branch. 

Future work requires a systematic search for correlated variability at
different orbital phases and as a function of X-ray energy, and to use
the information in order to constrain the binary parameters (mainly
inclination, binary separation and mass ratio) using appropriate
synthetic transfer functions \cite{teo05}.

\begin{theacknowledgments}
JC acknowledges support by the Spanish MCYT grant AYA2002-0036 and the
programme Ram\'on y Cajal.

\end{theacknowledgments}

\end{document}